\documentclass[11pt, letterpaper]{article}

\usepackage[letterpaper, margin=0.7in]{geometry}
\usepackage{amsmath,amssymb,amsfonts,amsthm}
\usepackage{algorithmic, enumerate}
\usepackage{graphicx}
\usepackage{textcomp}
\usepackage[numbers,square]{natbib}

\newtheorem{definition}{Definition}

\title{Inapplicability of the TVOR Method to \\ USHMM Data Outlier Identification}
\author{Melkior Ornik\thanks{Coordinated Science Laboratory and Department of Aerospace Engineering, University of Illinois Urbana-Champaign, Urbana, IL 61801 USA (e-mail: mornik@illinois.edu)}}
\date{}

\begin{document}

\maketitle

\begin{abstract}
Recent paper ``TVOR: Finding Discrete Total Variation Outliers Among Histograms'' introduces the Total Variation Outlier Recognizer (TVOR) method for identification of outliers among a given set of histograms{. After providing a theoretical discussion of the method and verifying its success on synthetic and population census data, it} applies the TVOR model to histograms of ages of Holocaust victims produced using United States Holocaust Memorial Museum data. It purports to identify the list of victims of the Jasenovac concentration camp as potentially suspicious. In this {comment} paper, we show that the TVOR model and its assumptions are grossly inapplicable to the considered dataset. When applied to the considered data, the model is biased in assigning a higher outlier score to histograms of larger sizes, the set of data points is extremely sparse around the point of interest, the dataset has not been reviewed to remove obvious data processing errors, and, contrary to the model requirements, the distributions of the victims' ages naturally vary significantly across victim lists.
\end{abstract}

{\small \noindent\textbf{Notice of publication}: This paper has been published in \textit{IEEE Access}, vol.~9, pp.~78586--78593, 2021, under the title ``Comment on 'TVOR: Finding Discrete Total Variation Outliers Among Histograms' ''. The difference in titles is due to the journal policy on naming of comment papers.}

\section{Introduction}
\label{intro}

Focusing on the problem of identifying compromised data, recently published article \cite{BanEle21} introduces a novel method named Total Variation Outlier Recognizer (TVOR) for identification of outliers across a set of histograms. In proposing its scheme based on the difference in discrete total variations among histograms, the TVOR method critically relies on the assumption that all histograms in a dataset should come from the same probability distribution, or should at least have the same smoothness properties.

{Initial experiments in \mbox{\cite{BanEle21}} focus on outlier detection in synthetic datasets drawn from a normal distribution, from a beta distribution, and from subsamples of the German census of 1939. TVOR method indeed successfully recognizes the outliers in those datasets.} Following these experiments, the work in \cite{BanEle21} considers data about Holocaust victims made available by the United States Holocaust Memorial Museum (USHMM) \cite{USH}. By comparing histograms obtained from {7106} historical documents such as lists of ghetto inhabitants, lists of casualties, census records, and concentration camp population lists --- including the lists of victims of the Jasenovac concentration camp \cite{USHJ} differentiated by ethnicity --- the authors of \cite{BanEle21} claim to have detected ``the potentially problematic parts of a sample, which in the case of the Jasenovac list lies in the birth years of Serbian inmates'' \cite[Appendix~D]{BanEle21}.

In this comment paper we show that the use of TVOR on the USHMM records, in the manner employed in \cite{BanEle21}, is inappropriate. We identify multiple features that make TVOR and its assumptions inapplicable to the USHMM dataset:
\begin{enumerate}[(i)]
\item {the histograms are not} drawn from the same probability distribution;
\item the model, when applied to the particular dataset, is biased towards providing a higher outlier score to larger lists such as the Jasenovac list; and
\item the dataset is sparse around the point of interest, and whatever data does exist does not satisfy the assumption of the histograms sharing similar {smoothness}, let alone the same probability distribution.
\end{enumerate}
We also show the method used in \cite{BanEle21} to pull the dataset from the USHMM database results in a significant contamination of the dataset, by duplicating hundreds of thousands of entries.

{We emphasize that we do not contest the theoretical underpinnings of TVOR method given in the first half of \mbox{\cite{BanEle21}}, and recognize the method to be valid in a variety of contexts. Our comment is solely on the application of the TVOR method to the particular dataset.}

\vskip 5pt

\noindent{\textbf{Notation.} In the remainder of the paper, we use the notation $\mathbb{N}_0$ for nonnegative integers and $[n]$ for set $\{1,\ldots,n\}$. Notation $\mathbb{E}[X]$ denotes the expected value of a random variable $X$ on an underlying probability space.}

\section{TVOR Preliminaries and USHMM Dataset}
\label{prel}

{Motivated by a classical problem of detection of outlying data \mbox{\cite{Cra46,Krietal09}},} TVOR \cite{BanEle21} seeks to identify outliers in a finite set of histograms $\{h_1,\ldots,h_k\}$.

\begin{definition}
A {\em histogram} is defined as a function $h:[n]\to\mathbb{N}_0$, where $h(i)$ signifies the amount of {\em samples} in a {\em bin} $i$. We refer to the sum $\sum_{i=1}^n h(i)$ as the {\em size} of the histogram.
\end{definition}

{The technical approach of TVOR relies on the fact that histograms drawn from the same probability distribution will, informally stated, share similar smoothness. To formally encode the notion of smoothness, \mbox{\cite{BanEle21}} proposes measuring a histogram's \textit{discrete total variation} ({DTV}).}

\begin{definition}
A {\em discrete total variation} ({DTV}) of a histogram $h:[n]\to\mathbb{N}_0$ is given by $\|h\|_V=\sum_{i=2}^n |h(i)-h(i-1)|$.
\end{definition}

{To detect outliers, TVOR then compares the DTV of each histogram with its expected value across the set ${h_1,\ldots,h_n}$, where the expectation is calculated by a best curve fit of the DTV data to a function of form $aN+b\sqrt{N}$.}

\begin{definition}
An {\em expected {DTV}} for a set of histograms $\{h_1,\ldots,h_k\}$, where $h_i:[n]\to\mathbb{N}_0$, is a function $$m(N)=aN+b\sqrt{N}$$ which best fits the set $\{(N_i,\|h_i\|_V)~|~i\in[k]\}$, where $N_i$ is the size of histogram $i$.
\end{definition}

While \cite{BanEle21} does not seem to explicitly state the metric of best curve fit for the expected {DTV}, the presented results indicate that the experiments consider a standard least squares fit or a similar metric. 

{Form $aN+b\sqrt{N}$ for the fitting function $m$ is motivated in \mbox{\cite{BanEle21}} by a bound of} 
\begin{equation}
\mathbb{E}[\|h\|_V]\leq \|\mathcal{D}\|_VN+\mathbb{E}[\|\mathcal{R}\|_V]\sqrt{N}
\label{boun}
\end{equation}
{derived in \mbox{\cite{BanEle21}}, where $N$ is the size of histogram $h$, $\|\mathcal{D}\|_V$ is the total variation of a distribution $\mathcal{D}:[n]\to[0,1]$ from which $h$ is drawn, and $E[\|\mathcal{R}\|_V]$ is the expected deviation between the histogram drawn from $\mathcal{D}$ and the distribution $\mathcal{D}$ itself. We invite the reader to consult \mbox{\cite[Sec.~III-D]{BanEle21}} for further details on the derivation of the bound. For the purposes of our paper, we emphasize, however, that values $\|\mathcal{D}\|_V$ and $E[\|\mathcal{R}\|_V]$ depend on the distribution from which $h$ is drawn. Hence, if multiple histograms are drawn from different distributions, the coefficients next to $N$ and $\sqrt{N}$ in \mbox{\eqref{boun}} may be different.}

{Finally, \mbox{\cite{BanEle21}} defines t}he TVOR outlier score $d'$ for a histogram $h:[n]\to\mathbb{N}_0$ of size $N$ belonging to the set $\{h_1,\ldots,h_k\}$ by normalizing the difference of its {DTV} with the expected {DTV} for a histogram of its size: \begin{equation}
\label{dd}
d'=\frac{|\|h\|_V-m(N)|}{\sqrt{N}}\textrm{.}
\end{equation}
{The normalization in \mbox{\eqref{dd}} is motivated in \mbox{\cite[Sec.~III-F]{BanEle21}} by the square root law \mbox{\cite{Fin09}}, i.e., the observation that the standard error of a sum of $N$ samples drawn from a distribution is often proportional to $\sqrt{N}$; for more details, we refer the reader to \mbox{\cite[Sec.~III-F]{BanEle21}}. We note that, if existent, the proportionality constant between the standard error and $\sqrt{N}$ is dependent on the distribution \mbox{\cite{Fin09}}; different distributions may lead to different constants.}

In parallel with TVOR, \cite{BanEle21} also proposes its modification \begin{equation}
\label{ddd}
d''=\frac{|\|h\|_V-\hat\mu_N|}{\hat\sigma_N}\textrm{,}
\end{equation} where instead of $m(N)$ and $\sqrt{N}$, the model considers the mean and standard deviation of {DTVs} of appropriately sized subsets of the German census of 1939 \cite{Sta40}.

{Experimental work in \mbox{\cite{BanEle21}} begins by successfully demonstrating TVOR on synthetic datasets and subsets of the German 1939 census data, which --- by design --- satisfy the assumption of shared probability distributions across histograms. The method performs well on these experiments, recognizing the outlying histograms successfully.}

The later part of \cite{BanEle21} focuses on the dataset compiled from the United States Holocaust Memorial Museum (USHMM) lists of Holocaust victims. The vast majority of the data used in that experiment is provided by the authors of \cite{BanEle21} online, in \cite{BanEle20}. All the data that we use in our experiments is contained in that repository. The full dataset consists of years of birth for {more than 3.6 million} individual records across {7106} lists. The lists are obtained from historical documents of varying provenance, including wedding announcements, ghetto inhabitant lists, death lists, concentration camp lists, and census records {\mbox{\cite{USH,Afo20}}}. The experiment in \cite{BanEle21} pays particular attention to the list of victims of the Jasenovac concentration camp, marked in the USHMM records by ID 45409 \cite{USHJ}. In addition to the full Jasenovac list, it also considers its multiple sublists, with victim records separated by ethnicity. These sublists do not seem to be available {in repository} \cite{BanEle20}, but they are described in detail in \cite{BanEle21}.

Each list in the dataset forms a natural histogram, where each bin corresponds to a particular year of birth. Some of the lists contain a small number of obviously impossible dates of birth, as discussed later; in our experiments, following the figures in \cite{BanEle21}, we a priori remove all years prior to $1850$ or after $1945$, resulting in $n=96$ bins {and a total of 3~619~428 records}.

The authors of \cite{BanEle21} then apply the TVOR method to identify the outlying histograms. They note that the Jasenovac list results in the highest $d'$-score, with $d'\approx 43.13$, thus labeling its data as ``potentially problematic'' {\mbox{\cite[Sec.~IV-F]{BanEle21}}}. After adding the ethnicity sublists within the Jasenovac list to the histogram set, the list of Serbian victims also yields a high $d'$-score of $d'\approx 40.82$ and is also described as potentially problematic {\mbox{\cite[Appendix~D]{BanEle21}}}. Similar $d''$ scores are obtained for Jasenovac from \eqref{ddd} by replacing the histogram set statistics with the German census of 1939.

\section{Inapplicability of Assumptions}
\label{nassum}

We claim that the application of TVOR to USHMM data does not in reality indicate any potential problems with the Jasenovac list or any of its subsets. In fact, TVOR is entirely inapplicable to the USHMM dataset as the dataset does not even come close to satisfying the fundamental assumption of TVOR --- that all lists in the set, except possible outliers, should share {similar probability distributions or at least DTVs}, and, in the case of the modified TVOR in \eqref{ddd}, also share similar DTVs to the German 1939 census. Such an assumption is acknowledged in \cite[Sec.~I]{BanEle21} and throughout \cite{BanEle21}.

In their only discussion of this challenge, the authors of \cite{BanEle21} write ``The geographical locations of these populations differ, but they still mostly cover the populations whose birth year histograms should have similar discrete total variation properties'' \cite[Sec.~IV-D2]{BanEle21}. {However, such a broad claim that birth year histograms should \textit{a priori} be expected to have similar DTVs is not supported in \mbox{\cite{BanEle21}}.} Such a claim is easily disproved {in general} by a wealth of historical evidence, as well as by data analysis.

The USHMM lists have been obtained from a variety of historical documents \cite{USH} of varying origin and purpose. For instance, the two largest lists in the collection (ID 25274 and 20619) deal with the inhabitants of the \L\'{o}d\'{z} Ghetto and contain population registry books kept by the ghetto's Judenrat (``Jewish Council'') \cite{USH25274,USH20619}. These books span the breadth of the Jewish community of the ghetto, from newborns to 90-year-olds \cite{USH20619}. On the other hand, list ID 20666 contains children sent from Theresienstadt to Auschwitz. The theoretical probability distribution generating the two groups cannot be the same: a list of children by definition cannot include an adult. {There is also no reason to believe that the total variations in different lists should be similar; to provide an explicit example, Sec.~\mbox{\ref{ndense}} illustrates variance in DTVs, and not solely probability distributions, showing that these --- unsurprisingly --- also differ significantly across 16 lists of similar size.}

\begin{figure}[h]
\centering
\includegraphics[width=\textwidth]{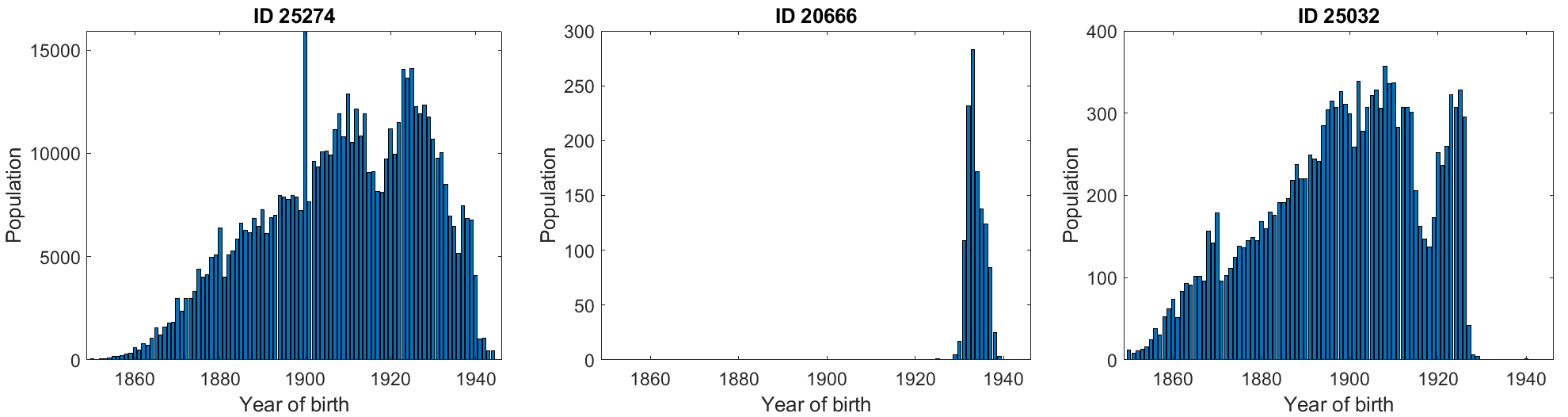}
\caption{Histograms of years of birth for population on lists ID 25274, 20666, and 25032. List ID 25274 spans a population across ages, but includes an obvious spike at 1900 due to the phenomenon of age heaping in unreliably collected data \cite{Rutetal90}. List ID 20666 includes only children. List ID 25032 excludes those ineligible for identification cards and thus cannot include almost any minors \cite{Cap13}.}
\label{fig1}
\end{figure}

We present the histograms compiled from lists ID 25274, 20666, and 25032 --- the latter describing holders of identification cards in Krakow \cite{USH25032} --- in Figure~\ref{fig1}. As expected, the distributions are significantly different. Such differences arise across the dataset. For instance, the Jasenovac list, by the nature of Jasenovac as a concentration camp for victims of multiple ethnicities and political prisoners \cite{BarDic17}, cannot be expected to share either the distribution, e.g., of the \L\'{o}d\'{z} Ghetto or of Theresienstadt children.

\section{Histogram Size Bias}
\label{nbias}

Having discussed the assumptions imposed on the dataset by applying the TVOR approach, we now move to show that{,} even if these assumptions were true for the considered dataset{,} the TVOR approach is biased towards assigning higher outlier scores to longer lists. {As described in Sec.~\mbox{\ref{prel}}, e}xactly to avoid size bias { --- an undesirable feature also noted in other statistical tests \mbox{\cite{Hab88}} --- } the {TVOR outlier score is normalized} by dividing the difference between the actual histogram DTV and the expected DTV by $\sqrt{N}$, where $N$ is the histogram size. {However, as noted in Sec.~\mbox{\ref{prel}}, the motivation for both this normalization and the choice of the class $aN+b\sqrt{N}$ for the expected DTV is reliant on all histograms being drawn \textit{from the same probability distribution}}. The authors of \mbox{\cite{BanEle21}} state that in practice such a normalization works well, ``even though it may introduce inaccuracies''. In the case of this dataset, {where underlying distributions, as well as their DTVs, are different,} such a normalization indeed does not completely remove bias.

To show the existence of bias, we {use several methods. We first calculate the correlation coefficient between histogram sizes and $d'$ scores. Its value is indeed positive and equals $0.17$. To place this value in comparison, the ``random noise'' correlation coefficient between the dataset histogram sizes and $7106$ samples chosen from a normal distribution is generally smaller by a factor of, roughly, $10$--$100$.}

{Motivated by the established correlation, we} use a linear regression model on the set $\{(N_i,d'(h_i)~|~i\leq 7106\}$ of USHMM histograms, where $N_i$ is the size of histogram $h_i$, and $d'(h_i)$ is its outlier score computed from \eqref{dd}. We easily find that the best fit is given by $$\overline{d}'(N)=4.017\cdot 10^{-5}N+2.128\textrm{.}$$ 
A similar fit is obtained by excluding the Jasenovac list from the fitting. While the coefficient of $4.017\cdot 10^{-5}$ may not seem significant for small $N$, we keep in mind that the Jasenovac list contains $N=78493$ elements. {We emphasize that we have no reason to believe that linear regression, or any other model, provides the best description for the relationship between sample size and $d'$-score. We simply first chose linear regression because of its simplicity and correspondence to correlation \mbox{\cite{WooEll07}}. We also certainly do not claim that the $d'$-score depends \textit{just} on the sample size.} 

Keeping in mind the identified bias, we now perform a small numerical experiment {to quantify its impact}. We renormalize $d'$ by considering $d_{ren}'=d'/\overline{d}'(N_i)$ to be the new outlier score; if the original scoring was unbiased, such a renormalization would have no impact. However, in this case it renders Jasenovac list's outlier score $d_{ren}'=8.15$ no longer the highest in the dataset. There are multiple histograms with a higher score, the highest of which ({$d_{ren}'=14.36$}) belongs to list ID 40168. A different regression model motivated by \cite{BanEle21}, given by $\overline{d}'(N)=aN+b\sqrt{N}+c$, also yields a renormalization in which the Jasenovac list does not feature prominently; after such a renormalization, the Jasenovac list has a lower outlier score than more than 50 other lists.

\begin{figure}[h]
\centering
\includegraphics[width=0.85\textwidth]{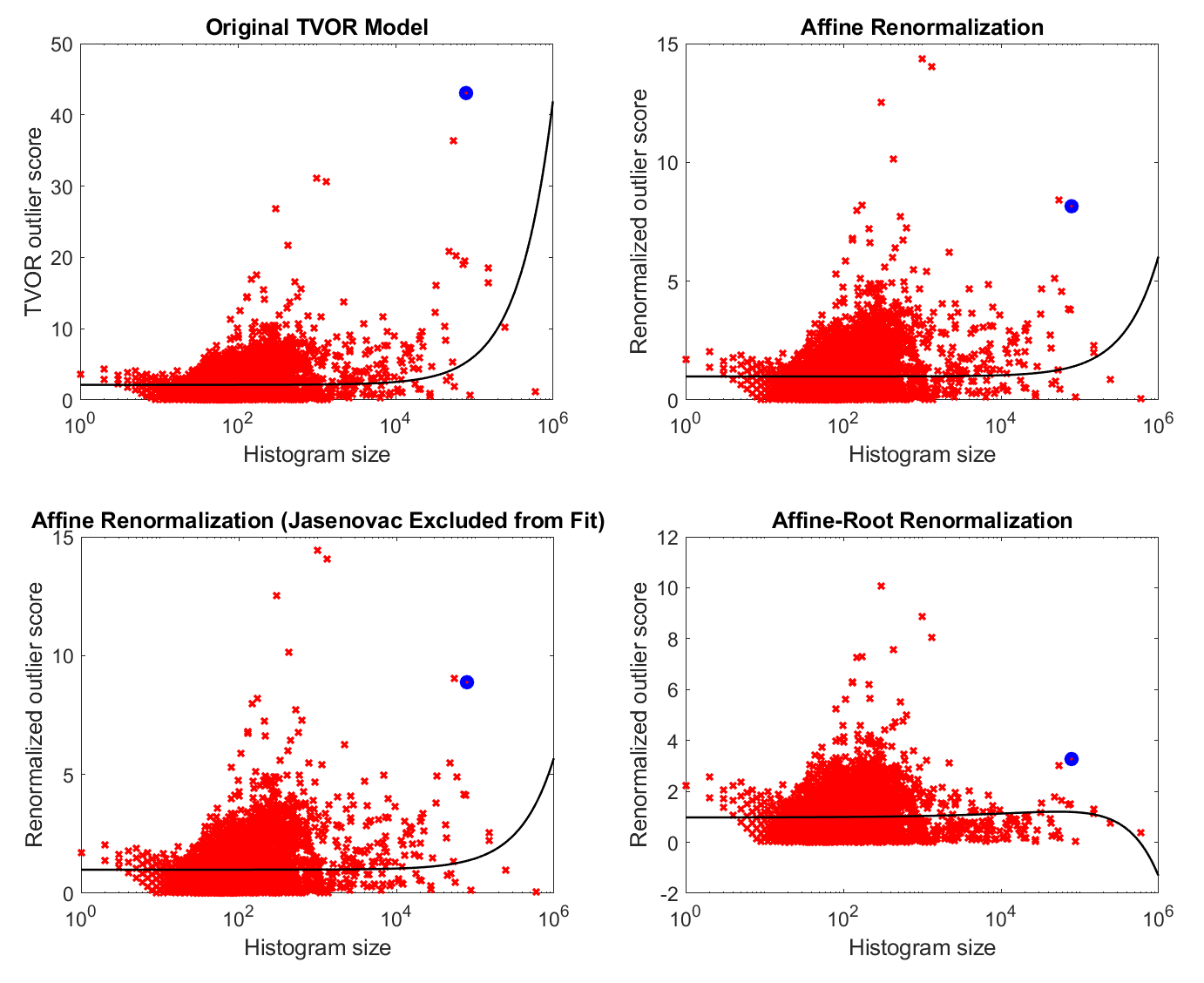}
\caption{A comparison between the results of the original TVOR model and the results of the three renormalized models: with the best fitting affine renormalization function $\hat{d}'$, with the best fitting affine renormalization function after the Jasenovac list is excluded from the dataset, and with the best fitting function of form $aN+b\sqrt{N}+c$. The scores for the Jasenovac list are marked in blue. The black line indicates the ``expected value'' of the outlier score for a histogram of a given size.}
\label{fig2}
\end{figure}

We present the resulting renormalized outlier scores in Figure~\ref{fig2}. While renormalization still does not entirely remove all size bias, it significantly decreases it --- an average histogram of size $10^6$ would get assigned $d_{ren}'<7$, in contrast to the original TVOR model which would assign $d'\approx 42$.

The identified bias is actually alluded to in \cite[Appendix~D]{BanEle21}, in the discussion of the Jasenovac sublists. After noting that even the sublist with the highest $d'$ score has a lower score than the entire list, the authors indeed attribute it to ``the fact that the sample with birth years of Serbian inmates has fewer values than the whole
Jasenovac list [...]. Because of that, such similar deviations are considered to
be less likely on a larger sample and thus the whole Jasenovac
list has a slightly larger value of score $d'$''. They, however, still proceed to claim that a high $d'$ score renders the Jasenovac list and the Serbian victims' sublist ``potentially problematic'' \cite[Appendix~D]{BanEle21}.

We emphasize that \textit{our discussion does not {purport} that there is anything potentially problematic about} list ID 40168 or any of the other lists that score higher than the Jasenovac list after any renormalization. {We emphasize that we are \textit{not} proposing any of the above models as a bona fide model for analysis of the USHMM data. We are solely showing that, given that the motivating assumptions for the TVOR model are not satisfied --- hence vitiating the method's theoretical foundation --- other similar models may produce different results. These models are less biased to size than TVOR in this particular case, but we make no claim that they are able to identify outliers in any way.} For all other reasons mentioned in this paper, the use of TVOR on the USHMM dataset cannot yield any meaningful conclusions even after {any} renormalization.

\section{Dearth of Relevant Histogram Data}
\label{ndense}

Let us now turn our attention to the third deficiency in applying TVOR, this time focusing on the specifics of applying it to the Jasenovac list. We will show that, \textit{even if} the assumptions made to use the TVOR method were applicable on the majority of the dataset and \textit{even if} the model was not biased, the lists of similar size to Jasenovac are few, and those that do exist have significantly different {DTVs} from each other, again in contravention of the assumptions in \cite{BanEle21}.

The vast majority of the USHMM dataset, as used in \cite{BanEle21}, consists of lists significantly smaller than the Jasenovac list. In fact, 74.5\% of lists have $\leq 100$ elements, more than $97\%$ have $\leq 1000$ elements, while more than $99\%$ have $\leq 10 000$ elements. We remind the reader that $N=78493$ in the Jasenovac list. There are only 6 lists (0.08\%) larger than the Jasenovac list, and --- as we will discuss later -- two of those are essentially the same (ID 20492 and ID 20493).

\begin{figure}[ht]
\centering
\includegraphics[width=0.5\textwidth]{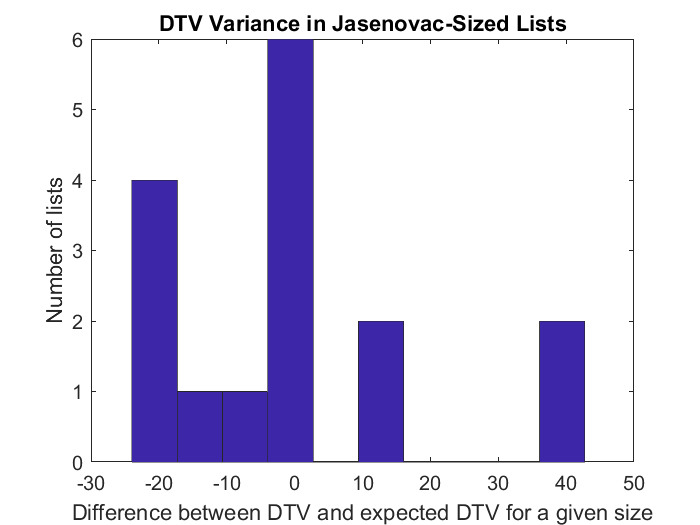}
\caption{The distribution of the difference between the DTV and expected DTV for lists similar in size to the Jasenovac list.}
\label{fig3}
\end{figure}

The few lists of similar size that do exist show divergent {DTV} properties, contradicting the assumption present in \cite[Sec.~III-A]{BanEle21} about similarities in smoothness (``[...] if this score is calculated for every sample in a group of samples that are expected to have similar smoothness, then the ones with the highest scores can be considered as outlier candidates''{, where \textit{smoothness} is somewhat circularly defined in the preceding sentence of \cite[Sec.~III-A]{BanEle21} as DTV}). If all lists, except perhaps Jasenovac, were to have similar smoothness, the signed $d'$ score $d'_\pm=(\|h\|_V-m(N))/\sqrt{N}$ would by definition be close to $0$ \textit{for all} those lists, where $m(N)$ is fitted as the expected DTV of those lists. However, when considering the dataset of the 16 lists similar in size to the Jasenovac list (with $39000\leq N\leq 157000$), only 6 lists have $d'_\pm$ score between $-5$ and $5$. Figure~\ref{fig3} illustrates the values of $d'_\pm$ for those 16 lists. The resulting distribution does not seem to have the vast majority of its mass around $0$.

\section{Data Processing and Analysis Issues}
\label{nerrors}

Having clarified the modeling issues, we now discuss the errors in the considered dataset. By consulting repository \cite{BanEle20} presented by the authors of \cite{BanEle21}, it becomes clear that the {authors'} method of downloading the USHMM data has led to substantial duplication.

Let us describe just several examples. The largest list in the dataset is list ID 25274, \textit{The Elders of the Jews in the \L\'{o}d\'{z} Ghetto}. However, that list is itself compiled from 14 440 sublists, some without accessible years of birth, which have their own IDs in the USHMM registry \cite{USH25274}. Nonetheless, \cite{BanEle21} considers these lists separately, in addition to the large list. {In fact, as can be readily checked by comparing the ID numbers in \mbox{\cite{BanEle20}} and \mbox{\cite{USH25274}}, 4780 lists in the dataset of \mbox{\cite{BanEle21}} are actually parts of the large \L\'{o}d\'{z} Ghetto list. This duplication amounts to \textit{67.2\% of all lists in the dataset}.}

{Among the remaining 33\% of the lists, there are additional obvious errors.} The authors of \cite{BanEle21} consider both lists ID 20492 and ID 20493 in their dataset, and even mention ID 20492 as the list with the third highest $d''$ score \cite[Sec.~IV-D3]{BanEle21}. However, those two lists are essentially the same \cite{USH20492,USH20493}. They both contain data from the registration cards of Jewish refugees in Tashkent. Their only difference, apart from three entries (out of 148 688) missing from one of the lists, is that one list is in the Latin alphabet, while the other is in Cyrillic. The dataset also considers lists ID 45680 and ID 45681 separately. However, when names not associated with a year of birth are removed, those two lists are entirely the same \cite{BanEle20}. {Naturally, the true extent --- and thus effect --- of the dataset contamination is likely impossible to measure given all the noted flaws in the dataset and the analysis of \mbox{\cite{BanEle21}}.}


We finally consider the element of the claims of \cite{BanEle21} that has little to do with statistics. As {corroboration} of their identification of potential problems with the Jasenovac list, \cite[Sec.~IV-F]{BanEle21} and \cite[Sec.~IV-G]{BanEle21} mention three of its inconsistent entries. The three entries describe individuals who are apparently likely to not have been in the Jasenovac camp. Without going into the merit of the listings for those three individuals, we first note that the dataset's Jasenovac list has more than 78~000 entries; it is reasonable to expect that any list of 78~000 people will have dozens of errors.

We then note that the authors of \cite{BanEle21} make no effort to identify or prune similar inconsistencies in other lists in the dataset, even when seemingly more egregious and more obvious. In fact, they use those lists directly in their algorithm. For instance, list ID 20652 recording deportations to Theresienstadt is claimed at its beginning to contain exactly 42155 persons \cite{USH20652}, but only 42150 are then listed. On the other hand, list ID 1394 is estimated on the USHMM website to contain 14 entries \cite{USH1394}, but it actually contains 61. List ID 25274, when converted in \cite{BanEle20}, seems to contain an individual born in 1771, who would have been around 170 years old at the time of record collection. \textit{We again emphasize that such small inconsistencies are to be expected in any project of the size of the USHMM records and do not indicate any widespread problem with the data}; we point them out only to indicate the lack of importance of the claim in \cite{BanEle21} about the three individuals on the Jasenovac list. 

Finally, we briefly comment on the discussion of \textit{age heaping} in \cite{BanEle21}. Indeed, age heaping --- a phenomenon where the year of birth of individuals whose age is uncertain is rounded{, often to $0$} --- is a common occurrence {\mbox{\cite{Rutetal90}}}. 
In Fig.~20 and Fig.~24--25 of \cite{BanEle21}, the authors draw {three classes} of claimed instances of age heaping in the Jasenovac list, a ``mid-decade pattern'' including the years ending in 4--8, {a heaping in years ending in 0,} and a heaping in years ending in 2. The paper performs no statistical analysis on these patterns. In fact, the mid-decade pattern is not described anywhere in \cite{BanEle21} other than being drawn on the figures. It is unclear how the authors of \cite{BanEle21} established the baselines for those years to identify heaping. The same is true for {the other two drawn classes of heaping}. {We emphasize that we make no claims about the actual occurrence of heaping or its quantitative properties, but simply note that the discussion on heaping in \mbox{\cite{BanEle21}} is not actually based on mathematical work presented in that paper.} The authors {of \mbox{\cite{BanEle21}}  also} do not mention a plausible historical explanation for the {seemingly more surprising} heaping in years ending in 2: regardless of their views on the total number of victims, a wide variety of sources agree that most victims in Jasenovac were likely killed in 1942 \cite{Ade04}. Thus, if the recording sources noted the victims' \textit{age} rather than \textit{year of birth}, age heaping will lead to a surplus of individuals seemingly born in a year ending with 2, by rounding the age at death to the nearest decade. We finally remark that the dataset in \cite{BanEle20} fails to take into account that some of the years of birth are marked as disputed on the USHMM website.

\section{Conclusion}

Paper \cite{BanEle21} introduces the Total Variation Outlier Recognizer (TVOR) method for recognizing anomalies in histogram sets and applies it{, among other datasets,} to lists from the United States Holocaust Memorial Museum {(USHMM)}, claiming to identify that the list of victims of the Jasenovac concentration camp --- and particularly the list of Serbian victims --- is potentially problematic.

In this paper, we show that{, unlike the synthetic datasets considered in earlier parts of \mbox{\cite{BanEle21}} which satisfy the method's assumptions,} the use of TVOR {on} the {USHMM} dataset is meaningless. {Sec.~\mbox{\ref{nassum}}--\mbox{\ref{ndense}} of our paper, respectively, show the following properties:}
\begin{itemize}
    \item the authors of \cite{BanEle21} disregarded the assumptions that they {used} in an earlier part of their paper, providing a dataset whose elements could not have possibly come from the same probability distribution,
    \item the normalization in the TVOR model to avoid a scoring bias to the size of each victim list is {inapplicable to the USHMM dataset}, and the model is still biased toward labeling larger lists as more outlying ones, and
    \item the number of lists comparable in size to the Jasenovac list is miniscule compared to the whole dataset, and data shows that even those comparable lists that do exist again do not satisfy {either} the assumption of {a shared distribution or a more relaxed assumption of} shared smoothness. 
\end{itemize}
Additionally, in {Sec.~\mbox{\ref{nerrors}}} we show that the data analyzed in \cite{BanEle21} was pulled from the USHMM database without necessary diligence, resulting in numerous partial or full duplications of the victim lists, contaminating the dataset. Finally, some of the purported patterns presented in \cite{BanEle21}, notably those on age heaping, were not obtained through the TVOR model at all, their mathematical provenance and statistical importance are unknown, and the authors of \cite{BanEle21} fail to identify simple plausible historical and data-collection explanations for them. 

In conclusion, while {other datasets analyzed in \mbox{\cite{BanEle21}} successfully show} the TVOR method may have its uses, attempted analysis of the Holocaust victim lists presented in \cite{BanEle21} is certainly not one of them.

\bibliographystyle{IEEEtran}
\small
\bibliography{refs}

\section*{Appendix}
\section*{Objections and Answers}

To clarify some of the important points of our paper, we use this appendix to specifically address some possible objections and concerns.

\vskip 5pt
\noindent \textit{Objection A: Are the models in Sec.~\ref{nbias} not less valid than TVOR, given that they lack TVOR's theoretical foundation?}

\vskip 5pt
\noindent \textbf{Answer: } On the dataset at hand, TVOR has no theoretical foundation. Its outlier scoring method is based on bound \eqref{boun}, which depends on the distribution from which the histograms are drawn, and the square root law \cite{Fin09}, where the proportionality constant between a standard error and $\sqrt{N}$ also depends on the histogram's theoretical distribution. As discussed in Sec.~\ref{nassum}, the histograms in the USHMM dataset are not drawn from the same distribution, and, as shown in Sec.~\ref{ndense}, the distributions from which they are drawn do not even have similar smoothness. The right hand side of \eqref{boun} thus yields differing coefficients next to $N$ and $\sqrt{N}$ across histograms. Analogously, the square root law provides differing proportionality constants across histograms, rendering both elements of TVOR's approach --- fitting to a function of form $aN+b\sqrt{N}$ and subsequent normalization through division by $\sqrt{N}$ --- theoretically groundless.

Unlike TVOR in \cite{BanEle21}, we make no claim in this paper that any of the models in Sec.~\ref{nbias} are valuable for outlier identification. In fact, we explicitly claim otherwise. The sole stated properties of these models are that (i) they suffer from less size bias than TVOR on the USHMM dataset and (ii) if TVOR had no size bias, they would produce the same results as TVOR. 

One could easily find many other models which yield either the same or differing results to TVOR or any of the other models from Sec.~\ref{nbias}. However, due to the results of Sec.~\ref{nassum} and Sec.~\ref{ndense}, any such model that assumes shared smoothness across histogram distributions cannot yield any meaningful results for the USHMM dataset.

\vskip 5pt
\noindent \textit{Objection B: Suppose that, even if one was to remove all duplicated and incorrect data, the outlier ranking still remains the same as in the analysis of USHMM data in \cite{BanEle21}. Would the approach of \cite{BanEle21} then be valid?}

\vskip 5pt
\noindent \textbf{Answer: } No. Firstly, we note that, while it is not difficult to remove the most egregious errors, it is near-impossible to remove \textit{all} duplicated and incorrect data as doing so would require access to historical sources that may no longer exist. For instance, it is a priori impossible to know whether two records with the same name and the same year of birth present a data processing duplicate or whether it happened that two individuals born in the same year share a name. Analogously, it is generally difficult to know whether a person recorded in lists of multiple concentration camps has been listed in one mistakenly or whether they were moved from one camp to another.

Secondly, even if all the errors in the dataset were removed, as discussed above, the TVOR method has no theoretical grounding on this dataset. Thus, there is no reason to suggest it would yield meaningful results even after corrections.

\vskip 5pt
\noindent \textit{Objection C: Suppose that, if one was to only consider lists of similar size and similar locations as the Jasenovac list (i.e., other concentration camps), the outlier ranking remains the same as in the analysis of USHMM data in \cite{BanEle21}. Would the approach of \cite{BanEle21} then be valid?}

\vskip 5pt

\noindent \textbf{Answer: } No. We first remark that, while only considering lists of similar size will alleviate the importance of size bias, doing so will reduce the dataset, making any statistical claims less impactful. For instance, as shown in Sec.~\ref{ndense}, the number of lists would be reduced by more than 99\% if one was to only consider lists with $N>10000$ (we note that $N=78493$ for the Jasenovac list).

However, assume for the sake of the argument that we do proceed with such a reduction. We show in Sec.~\ref{ndense} that even the 16 lists closest in size to the Jasenovac list do not share the same smoothness properties. Thus, an attempted application of TVOR would still lack theoretical foundation.

If one was to alternatively only consider lists of concentration camp victims as opposed to all USHMM lists, there is no reason to believe that those lists should share distributions or have similar smoothness. The reason is twofold. Firstly, camp populations were structurally different across concentration camps, owing to the camps' varying histories. For instance, the camp list closest in size to the Jasenovac one is the Flossenb\"{u}rg list (ID 44619). Flossenb\"{u}rg was a major source of forced labor throughout its existence, while Jasenovac had a significant child population \cite{BarDic17,Ade04}. Indeed, by consulting the dataset \cite{BanEle20}, only fewer than 1\% of persons on the Flossenb\"{u}rg list were recorded as having been born in 1930 or after. For Jasenovac, the proportion is around 22\%.

Secondly, the distributions of the lists and their smoothness do not only depend on actual camp populations but also on the method of list recording. Concentration camp lists have been collected in a variety of ways and are, as mentioned in Sec.~\ref{nerrors}, often incomplete or contain errors \cite{USH}. In short, an attempted application of TVOR to concentration camp data would also lack theoretical grounding due to inherent differences in underlying probability distributions.

\vskip 5pt

\noindent \textit{Objection D: There exists significant prior demographic work indicating that population distributions are often shared across regions and ethnicities. Is such work not in contradiction with the claims of Sec.~\ref{nassum}?}

\vskip 5pt

\noindent \textbf{Answer: } The vast majority of USHMM lists are not census records of populations in different areas. They are lists of people often grouped by age, status, political activity, or other factors. For instance, it is by definition impossible that a list of children, such as USHMM list ID 20666, will share the distribution of a list of identification card holders, such as list ID 25032. Sec.~\ref{ndense} consequently, and unsurprisingly, also shows that not only are population distributions provably different across lists in the dataset, but that they differ in smoothness.


\vskip 5pt

\noindent \textit{Objection E: Age heaping in the Jasenovac list is obvious, and is obviously problematic. Why does Sec.~\ref{nerrors} dispute it?}

\vskip 5pt

\noindent \textbf{Answer: } We first note that the notion of a feature being ``problematic'', while notably used in \cite{BanEle21}, has never been rigorously defined in that paper. It is not clear whether a well-known data recording artifact (such as age heaping) is to be considered problematic on its own, or whether there is an imputation of something more extreme or malicious.

That said, the discussion in Sec.~\ref{nerrors} states only that the comments on age heaping in \cite{BanEle21} are not founded on the TVOR method. None of Fig.~20, Fig.~24, or Fig.~25 from \cite{BanEle21}, which are used to discuss age heaping in that paper, have been obtained using TVOR.

\vskip 5pt

\noindent \textit{Objection F: Other statistical tests may also automatically assign a lower score to a smaller sample size. Does Sec.~\ref{ndense} thus imply that such statistical tests are generally worthless?}

\vskip 5pt

\noindent \textbf{Answer: } No. It solely suggests that different problems need to be carefully tackled by different tools. Namely, if a score produced by a statistical model depends on sample size, this raw score should then not be used across differently sized sample collections to make comparative claims about properties independent of sample size.

\end{document}